\documentstyle[12pt,epsfig]{article}
\begin{document}

Rapid Communications of Phys. Rev. {\bf D54}, R725 (1996)

\begin{flushright}
INR-0902/96 \\
January 1996 \\
hep-ph/9601335 \\
\end{flushright}

\begin{center}
{\large \bf New Effective Feynman-like Rules for
the Multi-Regge QCD Asymptotics
of Inclusive Multijet Production}\\
\vspace{0.3cm}
{\large  Victor T. Kim${}^{\dagger}$
and Grigorii B. Pivovarov${}^{\ddagger}$} \\
\vspace*{0.3cm}
{\em ${}^\dagger$
St.Petersburg
Nuclear Physics Institute,
 188350 Gatchina,
Russia}\footnote{\normalsize e-mail: $kim@pnpi.spb.ru$} \\
\vspace*{0.3cm}
{\em ${}^\ddagger$
Institute for Nuclear
Research, 117312 Moscow,
Russia}\footnote{\normalsize e-mail: $gbpivo@ms2.inr.ac.ru$}

\end{center}

\vspace*{1.0cm}
\begin{center}
{\large \bf
Abstract}
\end{center}
New effective Feynman-like rules are defined for inclusive multijet
cross sections in the multi-Regge regime.
The solution of the BFKL equation is used as a starting point.
The resulting rules involve conformal weight and
rapidity as a momentum and a coordinate respectively
and are translation invariant in the coordinates.
We use the effective rules to calculate ultra high energy
asymptotics of  inclusive multijet production.
The dependence on the parton densities
occurs only in the overall normalization of the asymptotic
cross sections. \\

\vspace*{2.0cm}
PACS number(s): 13.87.Ce,12.38.Cy,13.85.Hd

\newpage

The BFKL pomeron \cite{Lip76,Lip86} lies on the border
of the known territory
of perturbative QCD. It may give access to a new physics of
parton liquids \cite{Lev94} and it has inspired unexpected theoretical
developments
\cite{Kor95}, but it first requires experimental justification.
A natural place to look for that is in inclusive jet production in high
energy hadron collisions [5-15].
 The main problems
are: (i) to define selection criteria for the events that would ensure
the applicability of the BFKL analysis \cite{Ahn96}; (ii) to manage
the complexity of the BFKL asymptotics for the cross sections.
We deal here with the second problem. Namely, we give a new
representation for the asymptotic cross sections by means of effective
Feynman-like rules.
They factorize the full expressions for
the asymptotic cross sections into vertices and propagators of an
effective theory. A crucial property of the presented rules is
that they include propagators corresponding to the parton distributions.
That allows one to treat systematically integrations over parameters
of untagged most forward/bacward jets which are indispensable to 
calculations of
quantities measurable with the detectors of limited acceptance
\cite{Kim96}.

We illustrate the use of the
effective rules in a calculation of the ultra high energy asymptotic
behaviour of the cross sections.

We begin by defining kinematic regime. We consider the inclusive
differential
cross section of $N$-jet production at a fixed scale of large (with
respect to the hadronic scale) transverse momenta and very large
(much larger than the transverse momenta) total energy of colliding
hadrons.
The invariant mass of any jet pair is also supposed to be much
larger than the transverse momenta.
This is known as the multi-Regge regime.
As discussed in \cite{Kim96},
the cross section in this regime is a sum of terms each of which is,
crudely speaking,
a product of BFKL pomerons:
\begin{equation}
\label{sum}
d\sigma_N^{sum} =  d\sigma_N +
                     \int \left(d\Omega_f
                          \frac{d\sigma_{f+N}}{d\Omega_f}+
                          \frac{d\sigma_{N+b}}{d\Omega_b}
                                d\Omega_b\right)
                   + \int d\Omega_f
\frac{d\sigma_{f+N+b}}{d\Omega_fd\Omega_b}\Omega_b,
\end{equation}
where $d\sigma_N^{sum}$ is the differential cross section on
the phase space of $N$ massless partons
(jets for us are descendants of massless partons); $d\Omega_{f(b)}$
is the differential volume of the most forward (backward)
jet phase space;
and, for example, $d\sigma_{f+N}$ is the cross section on the phase
space of $N$ tagged jets plus the untagged most forward jet of an event.
The first term of the sum comes from events with the most forward and
backward jets both among the tagged jets (and it is absent in the case
of $N=1$).

We will describe simple rules to write down $d\sigma_N$
of Eq.(\ref{sum}).
To this end consider the simplest representative case of $N=3$.
$d\sigma_3$ integrated over the phase space of the most forward jet is
 (see Eq. (9) of Ref.\cite{Kim96})
\begin{eqnarray}
\label{bfklcomp}
\int d\Omega_f
     \frac{d\sigma_{f+2}}{d\Omega_f} = d\Omega_2
\frac{\alpha_{S} C_A}{k^2_{2\perp}}\frac{\alpha_{S} C_A}{k^2_{3\perp}}
\frac{2\alpha_{S} C_A}{\pi^2}\int_{x_2}^1 dx_1 F_A(x_1,\mu^2_1)
\int_{\mu_{1}}^{x_1 \sqrt{s}}\frac{d^{2}k_{1\perp}}{k_{1\perp}^2}
\times\nonumber\\
 \int d^{2}q_{\perp}f^{BFKL}(k_{1\perp},q_{\perp},y_1(x_1,k_{1\perp}))
f^{BFKL}(q_{\perp}+k_{2\perp},k_{3\perp},y)F_B(x_3,\mu^2_2),
\end{eqnarray}
where $d\Omega_2=dx_2dx_3d^2k_{2\perp}d^2k_{3\perp}$.

Let us explain the notations and meaning of Eq.(\ref{bfklcomp}).
The differential cross section of three jets integrated over the
phase space of the most forward jet is on the lhs of the equation.
The same integration in the rhs is over the longitudinal ($x_1$) and
transverse ($k_{1\perp}$) momenta of the most forward jet.
Longitudinal momenta are normalized to the half of the total
 energy and thus $x_1$
is the fraction of the hadron momentum carried
by the parton which scattered to produce the most forward jet.
As such it enters also as a variable of the effective parton
distribution function $F_A$ \cite{Com84} of the hadron $A$.
Another variable on which $F_{A,B}$ depend is the factorization scale
$\mu_{1,2}$ (one can take $\mu_{1,2}=\mu \sim min \{k_{i \perp}\}$).
$d\Omega_2$ is the differential volume
of the phase space of the two jets (all jets but the most forward
of the set $f+2$) while $x_i$ and $k_{i\perp}$ are their
longitudinal and transverse momenta respectively. The most remarkable
objects in the rhs are $f^{BFKL}$. They describe correlations
between transverse momenta of $t$-channel
reggeized gluons \cite{Lip76} emitted from a
pair of tagged jets nearest in rapidity space, and depend on
the rapidity intervals
\begin{eqnarray}
\label{rapidity}
y_1(x_1,k_{1\perp}) = \log \frac{x_1 k_{1\perp}}{x_2 k_{2\perp}}\,,
\nonumber\\
y = \log \frac{x_2x_3s}{k_{2\perp}k_{3\perp}}
\end{eqnarray}
spanned by the jet pairs (in the above formulae it is supposed
 that the middle  jet $2$ is in the forward direction; $s$ is
the squared total energy of the collision). The superscript
$BFKL$ is to recall that $f^{BFKL}$ is the solution of the
Balitsky-Fadin-Kuraev-Lipatov equation \cite{Lip76}. The
$f^{BFKL}$ depending on $y_1$ is called in Ref.\cite{Kim96} the
adjacent (to the hadron $A$) pomeron while that depending on $y$
is the inner pomeron (it is developed between the tagged jets).
The solution for the BFKL equation has the following integral
representation \cite{Lip76}:
\begin{equation}
f^{BFKL}(k_{1 \perp},k_{2 \perp},y)=
\sum_{n=-\infty}^{\infty}\int_{-\infty}^\infty d\nu
\chi_{n,\nu}(k_{1 \perp})e^{y\omega(n,\nu)}
\chi_{n,\nu}^*(k_{2 \perp}),
\label{lipatov}
\end{equation}
where the star means complex conjugation;
\begin{equation}
\chi_{n,\nu}(k_{\perp})=\frac{(k_{\perp}^2)^{-\frac{1}{2}+i\nu}
e^{in\varphi}}{2\pi}
\label{eigenfunction}
\end{equation}
are Lipatov's eigenfunctions and
$$\omega(n,\nu) = \frac{2 \alpha_{S} C_A}{\pi}
\biggl[ \psi(1) - Re \, \psi \biggl( \frac{|n|+1}{2} +
i\nu \biggr) \biggr]$$ are
Lipatov's eigenvalues. Here $\psi$ is the logarithmic derivative
of the Euler Gamma-function.
The summation in Eq.(\ref{lipatov}) runs over conformal spin indices $n$
and the integration is over conformal dimension $d = 1-2i\nu$.
Combinations $h=\frac{1+n}{2}-i\nu,\,\bar{h}=\frac{1-n}{2}-i\nu$
are known as conformal weights.

Here we should comment on the present status of our basic formulas
Eqs. (\ref{sum}) and (\ref{bfklcomp}). It is like the one of the formulas
of the naive parton model prior to the proofs of the QCD factorization
theorems (see for review Ref. \cite{Col89}). The recent  phenomenological
estimations of the applicability of the formulas like  
Eqs. (\ref{sum}) and (\ref{bfklcomp}) see, e.g., in Ref. \cite{Cud96}, 
and the attempt to prove the relevant factorization in Ref. \cite{Bal95}.

>From now on we start to restructure $d\sigma_{f+2}$ from
Eq.(\ref{bfklcomp}). First, one can integrate out the transverse
momentum $q_{\perp}$ of the $t$-channel reggeized gluon. To this end, the
following formula may be used:
\begin{eqnarray}
\label{trint}
\int \frac{d^2q_\perp}{q_\perp^2} \chi^*_{n,\nu}(q_\perp)
\chi_{m,\lambda}(q_\perp+k_\perp) = \pi \chi^*_{n,\nu}(k_\perp)
\chi_{m,\lambda}(k_\perp)\times
\nonumber\\
\frac{i^{|m-n|-|m|+|n|}}{\frac{|m-n|}{2}-i(\lambda-\nu+i\epsilon)}
\frac
{\Gamma\left(\frac{|m-n|}{2}+1-i(\lambda-\nu)\right)}
{\Gamma\left(\frac{|m-n|}{2}+1+i(\lambda-\nu)\right)}
\frac
{\Gamma\left(\frac{|m|+1}{2}+i\lambda\right)}
{\Gamma\left(\frac{|m|+1}{2}-i\lambda\right)}
\frac
{\Gamma\left(\frac{|n|+1}{2}-i\nu\right)}
{\Gamma\left(\frac{|n|+1}{2}+i\nu\right)}\,,
\end{eqnarray}
where $i\epsilon$ takes care of the singularity at
$m-n=\lambda-\nu=0$. The result of this transverse momentum
integration  is an integral representation for
$d\sigma_{f+2}$. Next, we rewrite it in new variables
for jet momenta and momenta of incoming hadrons $A$ and $B$.
To parametrize the light-cone components of hadron momenta
$p_A^+,\,p_B^-,\,s=p_A^+p_B^-$
we take
\begin{eqnarray}
x_{0}^+ = \log\frac{p_A^+}{\mu},
\nonumber\\
x_{4}^- = -\log\frac{p_B^-}{\mu},
\label{energy}
\end{eqnarray}
and to parametrize jet four-momenta  $k_i,\, i=1,2,3$
\begin{eqnarray}
\label{mom}
x^{+}_{i} = \log\frac{k_{i}^+}{\mu},
\nonumber\\
x^{-}_{i} = -\log\frac{k_{i}^-}{\mu},
\end{eqnarray}
where $k_{i\pm}=k_{i0}\pm k_{i3}$ are the light-cone components
of $k_i$ ($i=1$ corresponds to the most forward jet above).
A virtue of these variables is that the cross section
is invariant under translations $x_{i}^+\rightarrow x_{i}^+ + a,$
 $x_{i}^-\rightarrow x_{i}^- + a$:
\begin{eqnarray}
\label{xsect}
&\frac{sd\sigma_3}
{\pi^4\prod_{i=1}^{3}dx_{i}^+dx_{i}^-\frac{d\varphi_i}{2\pi}}=
\left(\frac{\alpha_SC_A}{2\pi^2}\right)^3
\sum_{n}\sum_{m}\int  d\nu\int d\lambda\, \nonumber\\
&\times
\left[G_A(x_{0}^+ - x_{1}^+;\mu)G(x_{1}^+ - x_{2}^+;-n,-\nu)
G(x_{2}^+ - x_{3}^+;-m,-\lambda)\right]
\nonumber\\
&\times
\left[U_{\varphi_1}(x_{1}^+-x_{1}^-;n,\nu)
R_{\varphi_2}(m-n,\lambda-\nu)
D_{\varphi_3}(x_{3}^+-x_{3}^-;-m,-\lambda)\right]
\nonumber\\
&\times
\left[G(x_{1}^- - x_{2}^-;n,\nu)G(x_{2}^- - x_{3}^-;m,\lambda)
G_B(x_{3}^- - x_{4}^-;\mu)\right],
\end{eqnarray}
where $d\sigma_3$ stands for $d\sigma_{f+2}$ of Eq.(\ref{bfklcomp});
$\varphi_i$ is the azimuthal angle of the $i$-th jet; and an explicit
form of the ``propagators'' $G_{A,B}$, $G$, $U$, $R$ and $D$ is
\begin{eqnarray}
\label{prop}
G_{A,B}(x;\mu)& = &\theta(x)F_{A,B}(e^{-x},\mu^2)\,,
\nonumber\\
G(x;n,\nu)& = &\theta(x)e^{-ix(\nu+i\frac{1+\omega(n,\nu)}{2})}\,,
\nonumber\\
U_\varphi(x;n,\nu)& = &\theta(x)i^{|n|}e^{in\varphi}
\frac
{\Gamma\left(\frac{|n|+1}{2}-i\nu\right)}
{\Gamma\left(\frac{|n|+1}{2}+i\nu\right)}\,,
\nonumber\\
R_\varphi(n,\nu)&=&\frac
{i^{|n|}e^{in\varphi}}
{\frac{|n|}{2}-i(\nu+i\epsilon)}
\frac
{\Gamma\left(\frac{|n|}{2}+1-i\nu\right)}
{\Gamma\left(\frac{|n|}{2}+1+i\nu\right)}\,,
\nonumber\\
D_\varphi(x;n,\nu)& = &(-1)^{|n|}U_\varphi(x;n,\nu).
\end{eqnarray}
Note the role of $\theta$-functions from Eq.(\ref{prop})
in Eq.(\ref{xsect}): they provide the right ordering
of the components of the hadron and jet momenta
($x_{A,i}^+$ and $x_{B,i}^-$ decrease
with an increment in $i$, and transverse momenta of the most
forward/backward jets are larger than the factorization scale
$\mu$);
the same ordering is seen in the
limits of integration over $x_1$ and $k_{1\perp}$
from Eq.(\ref{bfklcomp}).

We now consider the rhs of Eq.(\ref{xsect}) as corresponding
to a graph of Fig. 1. Namely, the first factor
$
\left(\frac{\alpha_SC_A}{2\pi^2}\right)^3
$
may be redistributed among the vertices of the graph;
summations over $n,m$ and integrations over $\nu,\lambda$
correspond to an integration over loop momenta. Each
momentum has a discrete ($n$ or $m$) and a continuous
($\nu$ or $\lambda$) component; the first square bracket
expression corresponds to the left hand side vertical line of the
graph and the last to the right one. Factors $U$ and $D$ of the
middle square bracket correspond to the up and down
border-rungs of the ladder graph respectively
and $R$ to the middle rung. Note also that the lines of the graph
are oriented and the sign of ``momentum'' variables of the
propagators depend on the direction of the momentum flow.

The next step is to note that one  obtains a more symmetric representation
for the Feynman-like rhs of Eq.(\ref{xsect}) if one
replaces loop momentum integrations by  equivalent integrations
over additional $x$- and $\varphi$- variables per vertex.
To this end, one multiplies the propagators of Eq.(\ref{prop})
by exponentials of products of the additional variables
and momenta in such a way that the additional
integrations provide momentum conservation at the vertices.
The momentum integrations may then be performed independently
for each ``propagator''; this will define the propagators
in the ``coordinate'' representation. In this way one arrives at
diagrams whose vertices are parametrized by two
$x$-variables and an azimuthal angle.
One may equally look at the resulting Feynman rules in the momentum
representation.
 Each momentum will consist of a  discrete
variable and two continuous variables.

We now describe the Feynman-like rules in the momentum
representation for the graph of Fig. 2 and then define
$d\sigma_N$ in terms of the analytic expression corresponding
to the graph.

Each vertex of the graph of Fig. 2 gives a factor
$\sqrt{\frac{\alpha_SC_A}{2\pi^2}}$.
Each momentum comprises two continuous and one discrete
variables (for example,  $k^{A,i} = (k_1^{A,i},k_2^{A,i},n^{A,i})$);
Momenta flowing along the arrows are calculated
with momentum conservation at the vertices as linear combinations
of the external and the loop momenta.
There are lines of six types: $G_{A}$, $G_B$, $G$, $U$,
$D$ and $R$.
Each line gives the following factor depending on its momentum and,
for the ladder rungs, on the azimuthal angles
$\varphi_i$ of the corresponding jets:

\begin{eqnarray}
\label{mprop}
G_{A,B}(k)&=&g_{A,B}(k_1),
\label{G_A}\\
G(k)&=&\frac{1}{2\pi i}\frac
{1}
{k_2-k_1+i\frac{1+\omega(n,k_2)}{2}-i\epsilon},
\label{G}\\
U_\varphi(k)&=&\frac{1}{2\pi i}\frac
{-1}
{k_1+i\epsilon}e^{in\varphi+iu(n,k_2)},
\label{U}\\
D_\varphi(k)&=&\frac{1}{2\pi i}\frac
{-1}
{k_1+i\epsilon}e^{in\varphi+id(n,k_2)},
\label{D}\\
R_\varphi(k)&=&\delta(k_1)\frac
{1}
{\frac{|n|}{2}-i(k_2+i\epsilon)}e^{in\varphi+ir(n,k_2)}\,,
\label{R}
\end{eqnarray}
where
\begin{eqnarray}
g_{A,B}(k)&=&\int\frac{dx}{2\pi}e^{ikx}\theta(x)F_{A,B}(e^{-x},\mu^2)\,,
\label{g}\\
iu(n,k)&=&i\frac{\pi}{2}|n|+\log
\frac
{\Gamma\left(\frac{|n|+1}{2}-ik\right)}
{\Gamma\left(\frac{|n|+1}{2}+ik\right)}\,,
\label{u}\\
id(n,k)&=&-i\frac{\pi}{2}|n|+\log
\frac
{\Gamma\left(\frac{|n|+1}{2}-ik\right)}
{\Gamma\left(\frac{|n|+1}{2}+ik\right)}\,,
\label{d}\\
ir(n,k)&=&i\frac{\pi}{2}|n|+\log
\frac
{\Gamma\left(\frac{|n|}{2}+1-ik\right)}
{\Gamma\left(\frac{|n|}{2}+1+ik\right)}\,.
\label{r}
\end{eqnarray}
The product is integrated over the loop momenta with the measure
\begin{equation}
\label{measure}
\sum_{n=-\infty}^\infty\int_{-\infty}^\infty dl_1
\int_{-\infty}^\infty dl_2.
\end{equation}
The result of the integration is multiplied by
\begin{equation}
\label{delta}
\delta(\sum_{i=0}^Nk^{A,i}_1-\sum_{i=0}^Nk^{B,i}_1).
\end{equation}
The final result is a function that we will denote as
\begin{equation}
\label{F}
I_N(k^{A,0},\dots,k^{A,N};k^{B,1},\dots,k^{B,N+1};\varphi_1,\dots,\varphi_N)=
I_N(k^{A};k^{B};\varphi).
\end{equation}
This completes the description of the Feynman rules.

The cross section in terms of $I_N(k^A,k^B,\varphi)$ is
\begin{equation}
\label{res}
\frac{sd\sigma_N}
{\pi^4\prod_{i=1}^Ndx_{i}^+dx_{i}^-\frac{d\varphi_i}{2\pi}}=
\int \left(\prod_{i=0}^N dk^{A,i}_1 e^{-ik^{A,i}_1x^+_i}\right)
\left(\prod_{i=1}^{N+1} dk^{B,i}_1 e^{ik^{B,i}_1x^-_i}\right)
I_N^0(k^A;k^B;\varphi)\,,
\end{equation}
where $I_N^0$ is $I_N$ at $k_2^{A,i}=k_2^{B,i}=n^{A,i}=n^{B,i}=0$
and the energy variable of the collision $s$ is connected
with $x^+_0,x^-_{N+1}$
by $s=\mu^2 e^{x^+_0-x^-_{N+1}}$ (see Eq.(\ref{energy})).

To illustrate the use of the above  rules let
us calculate asymptotic inclusive
single-jet cross section at high energy.
 The leading contribution comes from
events with the untagged most forward and most backward jets:
\begin{equation}
\label{sjl}
\frac{sd\sigma^{sum}_1}{\pi^4dx^+dx^-d\frac{\varphi}{2\pi}}\approx
\int dx^+_fdx^-_f\frac{d\varphi_f}{2\pi}
\frac{sd\sigma_3}{\pi^4dx^+_fdx^-_fd\frac{\varphi_f}{2\pi}
     dx^+dx^-\frac{d\varphi}{2\pi}
     dx^+_bdx^-_b\frac{d\varphi_b}{2\pi}}
     dx^+_bdx^-_b\frac{d\varphi_b}{2\pi}.
\end{equation}
Then the use of the above rules gives (see Fig 3.)
\begin{eqnarray}
\label{use}
&\frac{sd\sigma^{sum}_1}{\pi^4dx^+dx^-d\frac{\varphi}{2\pi}}\approx
\left(\frac{\alpha_SC_A}{2\pi^2}\right)^3
\frac{i}{(2\pi)^6}\int dk^Adq^Ae^{-ik^Ax^+_A-iq^Ax^+}
dk^Bdq^Be^{ik^Bx^-_B+iq^Bx^-}d^2l^Ad^2l^B\times\nonumber\\
&\left[g_A(k^A)\frac{1}
{(l^A_1-l^A_2-k^A+i\frac{1+\omega(0,l^A_2)}{2}-i\epsilon)
(l^A_2-l^A_1+i\frac{1+\omega(0,l^A_2)}{2}-i\epsilon)}\right]
\left[A\rightarrow B\right]
\times
\nonumber\\
&\frac{\exp\left(
iu(0,l^A_2)+id(0,l^B_2)+ir(0,-l^A_2-l^B_2)\right)}
{(l^A_1+i\epsilon)(l^B_1+i\epsilon)(l^A_2+l^B_2-i\epsilon)}
\times\nonumber\\
&\delta(k^A+q^A-l^A_1-l^B_1)\delta(k^A+q^A-k^B-q^B)
\,,
\end{eqnarray}
where the second square bracket expression is obtained from the first one
by the substitution $A\rightarrow B$ and
we took into account the fact that $\omega(0,k)$ is an even function.
 As $s=\mu^2 e^{x^+_A-x^-_B}$,
we are interested in the limit $ x^+_A\rightarrow\infty$,
$x^-_B\rightarrow-\infty$. To calculate it,
we first integrate out $q^A,q^B$ by means of the $\delta$-functions,
then take the residues at $k^A=l^A_1-l^A_2+i\frac{1+\omega(0,l^A_2)}{2}$,
$k^B=l^B_1-l^B_2+i\frac{1+\omega(0,l^B_2)}{2}$ (only these poles contribute
to the asymptotic limit),
then at $l^A_1=l^A_2+i\frac{1+\omega(0,l^A_2)}{2}$,
$l^B_1=l^B_2+i\frac{1+\omega(0,l^B_2)}{2}$, and finally take the
remaining
integrations over $l^A_2,l^B_2$ in the saddle point approximation
(the saddle point is $l^A_2=l^B_2=0$). The net result is
\begin{equation}
\label{asymptotic}
\frac{sd\sigma^{sum}_1}{\pi^4dx^+dx^-d\frac{\varphi}{2\pi}}\approx
\left(\frac{\alpha_SC_A}{2\pi^2}\right)^3
\frac{
e^{\alpha_{I \!\! P}(x^+_A-x^-_B)}
M_A({\alpha_{I \!\! P}},\mu^2)
M_B({\alpha_{I \!\! P}},\mu^2)}
{ (2 \pi \alpha_{I \!\! P})^2
\sqrt{14\alpha_S C_A\zeta(3)(x^+_A-x^-_B)} }\,,
\end{equation}
where
\begin{equation}
\nonumber
\alpha_{I \!\! P}=1+4 \frac{\alpha_S C_A}{\pi}\log 2
\end{equation}
is the BFKL pomeron intercept \cite{Lip86} and
\begin{equation}
\label{moment}
M_{A,B}(\alpha_{I \!\! P},\mu^2)=\int_0^1 dx
x^{\alpha_{I \!\! P}-1}
F_{A,B}(x,\mu^2)
\end{equation}
are moments of the parton distribution functions.

Note that the asymptotic cross section is independent of the
jet parameters and depends on the parton distribution functions
only by an overall normalization factor.
One may assess the usefulness of the above representation
of the cross sections trying to reproduce without it the asymptotic
of Eq.(\ref{asymptotic}) by integration over parameters
of most forward/backward jets
of the corresponding cross sections from Refs. \cite{Rys80,Del93} 
where the single jet production was considered under fixed
parameters of most forward/backward jets. This integration
changes the dependence of the asymptotic cross section of 
Refs. \cite{Rys80,Del93} on the parameters of the tagged jet.

The moments of Eq.(\ref{moment}) will enter also the asymptotics
of the inclusive multijet cross section. This may be obtained along the
same lines as the single-jet asymptotic limit of Eq.(\ref{asymptotic}).
We will present this elsewhere.

 The use of the above rules for the inclusive dijet production
reproduces the results of Ref. \cite{Kim96}. In particular, one
may obtain a diagrammatic representation for the BFKL structure functions
of Ref. \cite{Kim96}.
 
We would like to stress that presented  effective Feynman-like rules 
for description of inclusive cross sections
are complimentary to the effective field theory 
of interacting reggeized and physical gluons (Ref. \cite{Lip95}) which 
describes ''exclusive'' processes.  
However, exact relation of our effective rules and effective theory of Ref. 
\cite{Lip95} requires further study.

To sum up, we introduced new effective Feynman-like rules
for inclusive multijet cross sections in the multi-Regge regime,
and used them to calculate an ultra high energy asymptotic limit
of single jet production.

 We thank I.F. Ginzburg, L.N. Lipatov, A.H. Mueller, V.A. Rubakov and
 A.A. Vorobyov for stimulating discussions and comments.
 V.T.K. is indebted to M.G. Albrow and T. Lee for valuable discussions.
This work was supported in part by the Russian Foundation for Basic
Research, Grant No. 96-02-16717.

\vspace*{2cm}
{\large \bf Figure Captions}
 
\vspace*{1cm}
Fig. 1: Diagrammatic representation
of 3-jet cross section $\sigma_3$.

\vspace*{1cm} 
Fig. 2: Graph corresponding to the N-jet cross section $\sigma_N $.

\vspace*{1cm} 
Fig. 3: The graph giving leading contribution
to the asymptotic inclusive single-jet production cross section.

\newpage
\vspace*{5cm}
\begin{figure}[htb]
\includegraphics{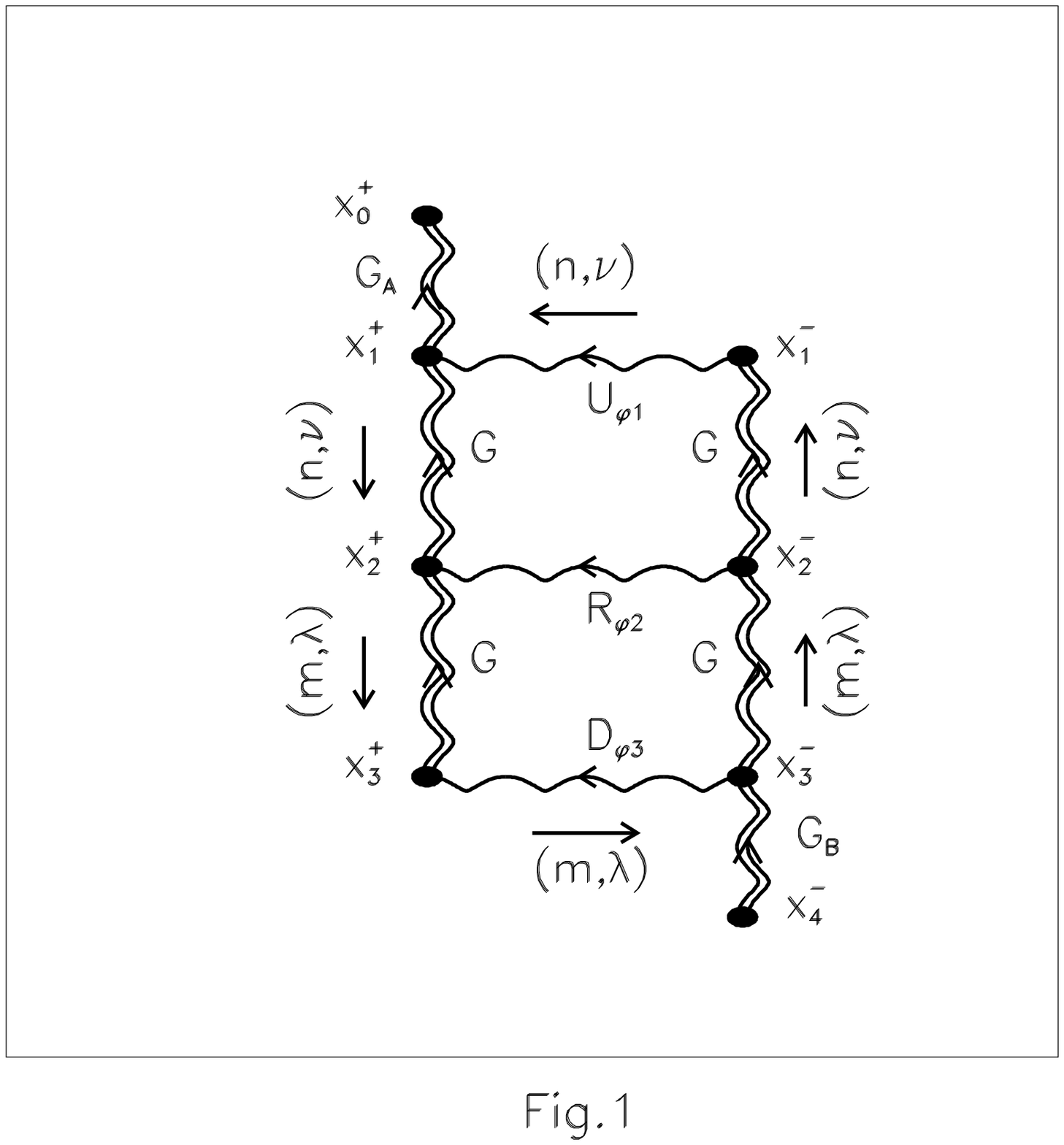}
\end{figure}

\newpage
\vspace*{5cm}
\begin{figure}[htb]
\includegraphics{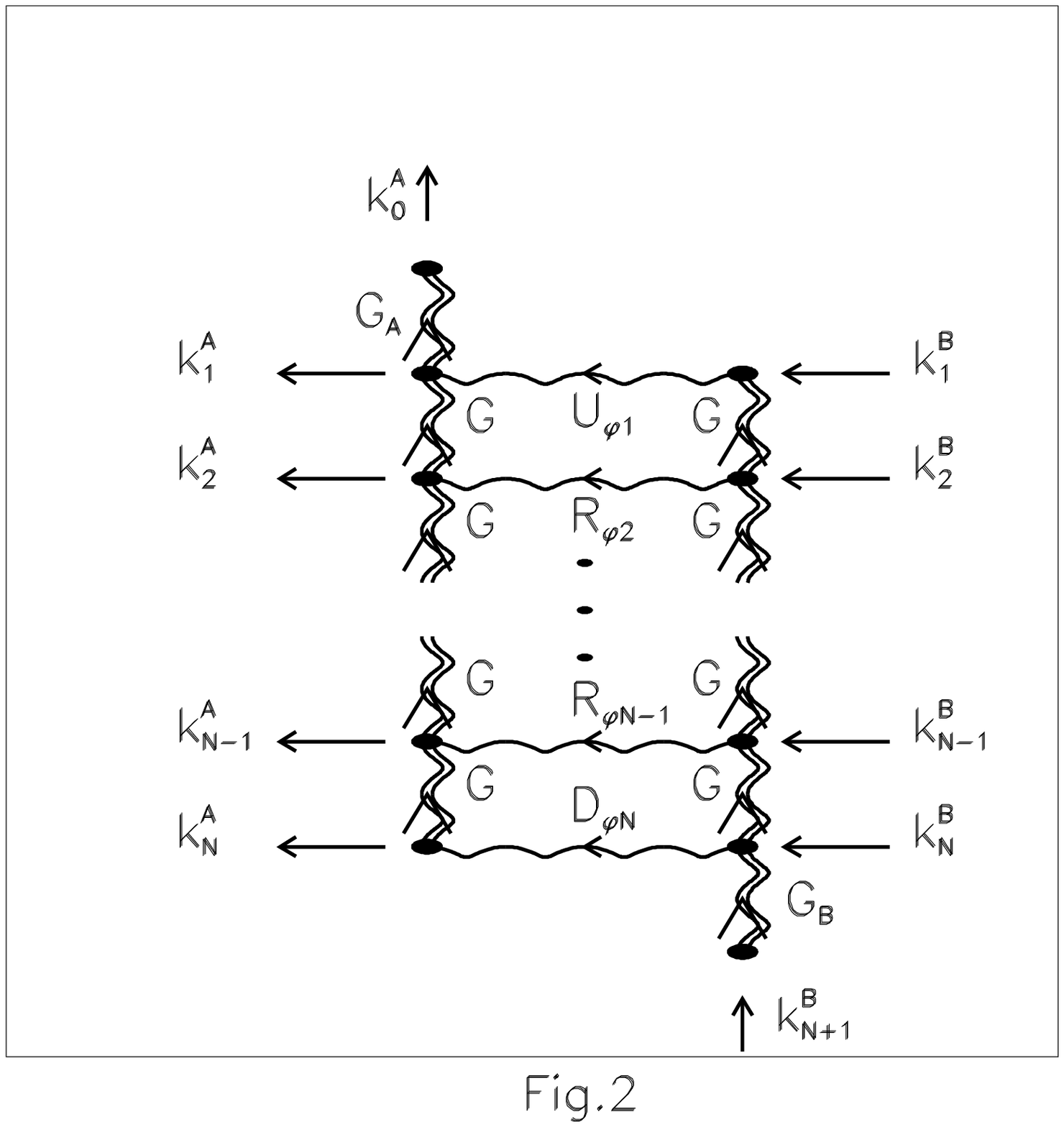}
\end{figure}

\newpage
\vspace*{5cm}
\begin{figure}[htb]
\includegraphics{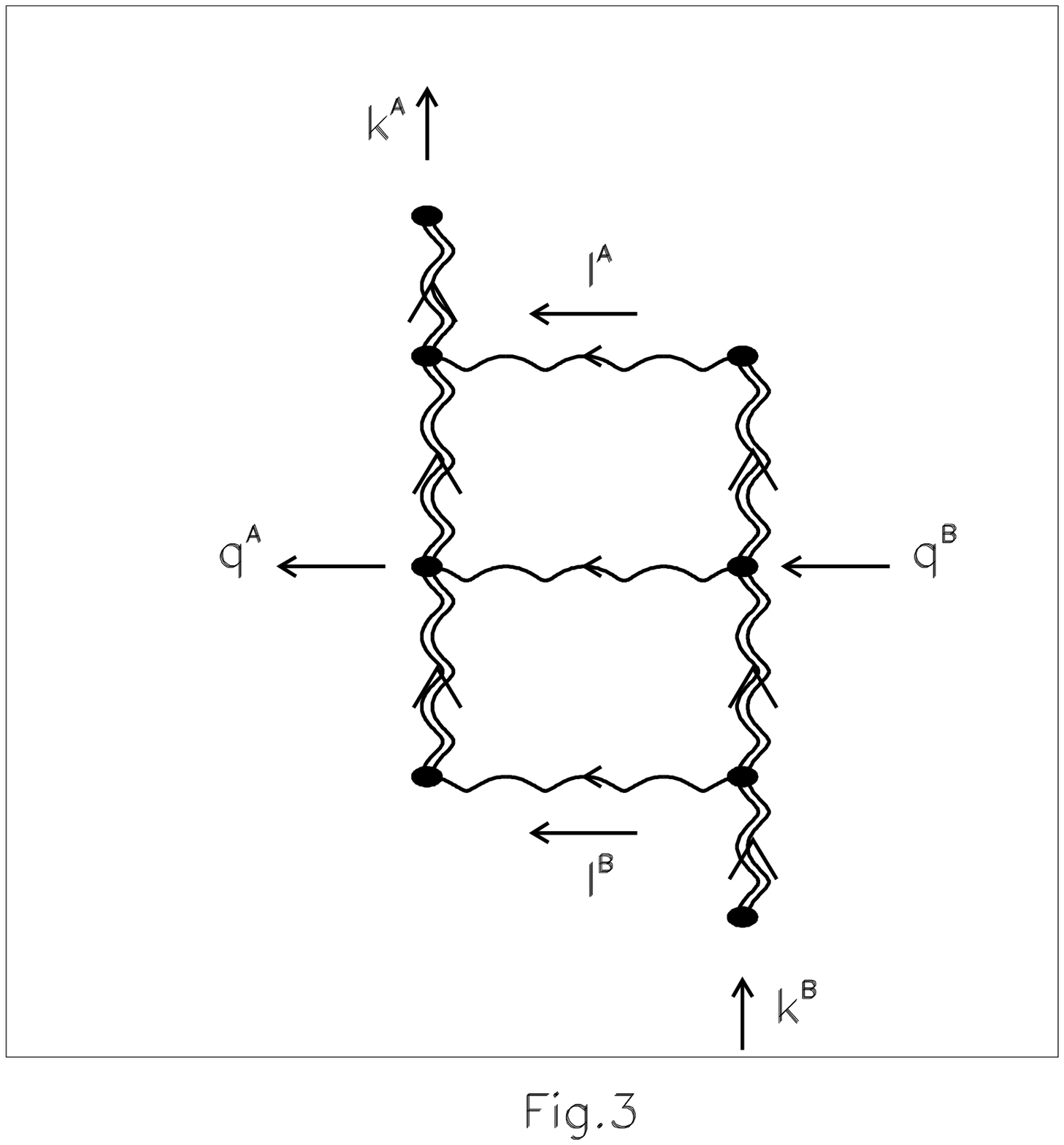}
\end{figure}

\end{document}